\documentclass[reprint,amssymb,amsmath,aps,superscriptaddress,showpacs,pra,footinbib,longbibliography]{revtex4-1}
\usepackage{graphicx}

\usepackage{color}
\usepackage[colorlinks,bookmarks=false,citecolor=darkblue,linkcolor=red,urlcolor=blue]{hyperref}
\def\cred{}
\definecolor{darkred}{rgb}{0.7,0.0,0.0}

\definecolor{darkblue}{rgb}{0,0.02,0.45}

\definecolor{darkgreen}{rgb}{0.02,0.45,0.0}

\definecolor{violet}{rgb}{0.8,0.2,0.6}

\newcommand{\eff}{\rm eff}

\begin{document}

\title{Cubic symmetry and magnetic frustration on the $fcc$ spin lattice in K$_2$IrCl$_6$}

\author{Nazir Khan}
\affiliation{Experimental Physics VI, Center for Electronic Correlations and Magnetism, Institute of Physics, University of Augsburg, 86135 Augsburg, Germany}

\author{Danil Prishchenko}
\affiliation{Experimental Physics VI, Center for Electronic Correlations and Magnetism, Institute of Physics, University of Augsburg, 86135 Augsburg, Germany}
\affiliation{Ural Federal University, Mira Str. 19, 620002 Ekaterinburg, Russia}

\author{Yurii Skourski}
\affiliation{Hochfeld-Magnetlabor Dresden (HLD-EMFL), Helmholtz-Zentrum Dresden-Rossendorf, D-01314 Dresden, Germany}

\author{Vladimir G. Mazurenko}
\affiliation{Ural Federal University, Mira Str. 19, 620002 Ekaterinburg, Russia}

\author{Alexander~A. Tsirlin}
\email{altsirlin@gmail.com}
\affiliation{Experimental Physics VI, Center for Electronic Correlations and Magnetism, Institute of Physics, University of Augsburg, 86135 Augsburg, Germany}
\affiliation{Ural Federal University, Mira Str. 19, 620002 Ekaterinburg, Russia}

\begin{abstract}
Cubic crystal structure and regular octahedral environment of Ir$^{4+}$ render antifluorite-type K$_2$IrCl$_6$ a {\cred model fcc antiferromagnet with a combination of Heisenberg and Kitaev exchange interactions}. High-resolution synchrotron powder diffraction confirms cubic symmetry down to at least 20\,K, with a low-energy rotary mode gradually suppressed upon cooling. Using thermodynamic and transport measurements, we estimate the activation energy of $\Delta\simeq 0.7$\,eV for charge transport, the antiferromagnetic Curie-Weiss temperature of $\theta_{\rm CW}\simeq -43$\,K, and the extrapolated saturation field of $H_s\simeq 87$\,T. All these parameters are well reproduced \textit{ab initio} using $U_{\rm eff}=2.2$\,eV as the effective Coulomb repulsion parameter. The antiferromagnetic Kitaev exchange term of $K\simeq 5$\,K is about one half of the Heisenberg term $J\simeq 13$\,K. While this combination removes a large part of the classical ground-state degeneracy, the selection of the unique magnetic ground state additionally requires a weak second-neighbor exchange coupling $J_2\simeq 0.2$\,K. {\cred Our results suggest that K$_2$IrCl$_6$ may offer the best possible cubic conditions for Ir$^{4+}$} and demonstrates the interplay of geometrical and exchange frustration in a high-symmetry setting.
\end{abstract}

\maketitle

\section{Introduction}
Strong spin-orbit coupling is an essential ingredient of correlated insulators with $5d$ transition metals, such as Ir$^{4+}$ that typically features an octahedral oxygen coordination along with five electrons and one hole in the $t_{2g}$ shell. In the absence of any additional crystal-field splitting, spin-orbit coupling separates the $t_{2g}$ states into the lower-lying $j_{\eff}=\frac32$ and higher-lying $j_{\eff}=\frac12$ manifolds, with the latter forming a half-filled band gapped by even moderate electronic correlations~\cite{rau2016}. This general scenario has been exemplified in more than a dozen of iridates studied over the last decade~\cite{cao2018}, although the symmetry of Ir$^{4+}$ is usually lower than cubic, thus leading to crystal-field splittings within the $t_{2g}$ shell~\cite{calder2014,rossi2017} or, in cases like CaIrO$_3$~\cite{moretti2014,kim2015} and Sr$_3$CuIrO$_6$~\cite{liu2012}, even to profound deviations from the $j_{\rm eff}=\frac12$ scenario. 

The quest for cubic systems based on Ir$^{4+}$ is triggered by interesting predictions for the magnetic interactions that would arise in this setting~\cite{rau2016,winter2017}. It has been proposed that the combination of the \mbox{$j_{\eff}=\frac12$} state and $90^{\circ}$ Ir--O--Ir superexchange leads to the bond-directional (Kitaev) anisotropy of exchange interactions~\cite{jackeli2009}, which, in turn, has broad implications for exotic quantum states and even topological quantum computing~\cite{kitaev2006}. This physics is presently explored in the honeycomb iridates A$_2$IrO$_3$ (A = Li, Na) and related materials~\cite{winter2017,hermanns2018}.

Here, we report on the crystal and electronic structures as well as the magnetic behavior of K$_2$IrCl$_6$, an Ir$^{4+}$ compound that retains its cubic symmetry and, thus, the ideal octahedral coordination of Ir$^{4+}$ down to low temperatures. This renders K$_2$IrCl$_6$ an interesting model material that combines the geometrically frustrated (fcc) arrangement of the Ir$^{4+}$ ions with sizable exchange anisotropy, a rare case among the $5d$ materials. We confirm the frustrated nature of K$_2$IrCl$_6$ experimentally, derive the relevant microscopic parameters, and discuss the extent of exchange anisotropy in this compound.

\begin{figure}
\includegraphics[width=7.5cm]{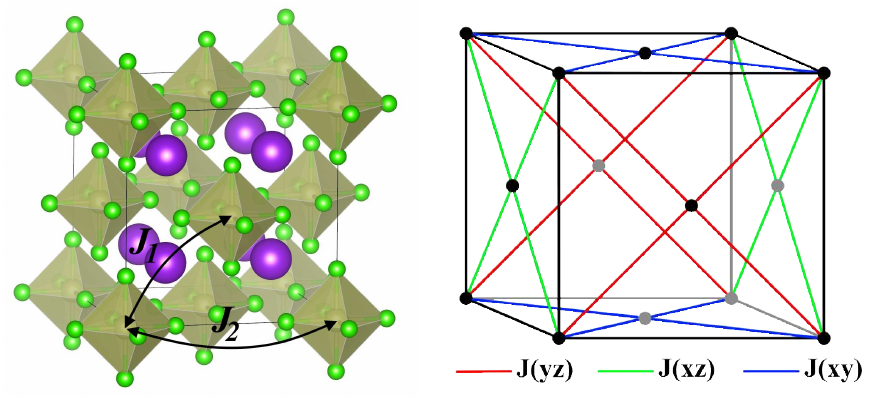}
\caption{\label{fig:structure}
Left: cubic crystal structure of K$_2$IrCl$_6$ and magnetic interactions therein. \texttt{VESTA} software~\cite{vesta} was used for visualization. Right: fcc spin lattice with different anisotropic interactions of the $J-K-\Gamma$ model on different faces of the cube (see Fig.~\ref{sec:model}).
}
\end{figure}

K$_2$IrCl$_6$ belongs to the K$_2$PtCl$_6$ family of cubic antifluorite-type A$_2$MX$_6$ hexahalides that feature isolated MX$_6$ octahedra arranged on the fcc lattice and separated by alkali-metal cations (Fig.~\ref{fig:structure}, left)~\cite{armstrong1980}. The magnetic behavior of K$_2$IrCl$_6$ was reported back in 1950's~\cite{cooke1959,griffiths1959}, but, surprisingly, even the exact crystal structure was not determined, and no microscopic information for this compound is available to date. Thermodynamic measurements~\cite{cooke1959,bailey1959,willemsen1977,moses1979} and neutron diffraction~\cite{hutchings1967,minkiewicz1968,lynn1976} suggest the onset of magnetic order below $T_N\simeq 3$\,K with the collinear spin structure and a possible field-induced magnetic transition~\cite{meschke2001}.

\section{Methods}
Powder samples of K$_2$IrCl$_6$ are commercially available from Alfa Aesar (Ir 39\% min) and were used without further purification. Sample quality was confirmed by powder x-ray diffraction (XRD) data collected at the Rigaku MiniFlex diffractometer (CuK$_{\alpha}$ radiation). High-resolution XRD was performed at the MSPD beamline~\cite{fauth2013} of the ALBA synchrotron facility ($\lambda=0.4129$\,\r A) and at the ID22 beamline of the ESRF ($\lambda=0.35456$\,\r A). The sample was placed into a thin-wall borosilicate capillary and cooled using the He cryostat. The capillary was spun during the data collection. Diffracted signal was recorded by 14 (ALBA) and 9 (ESRF) point detectors preceded by Si (111) analyzer crystals. \texttt{Jana2006} software was used for the structure refinement~\cite{jana2006}. No crystalline impurity phases were detected in the commercial samples from Alfa Aesar within the sensitivity of the synchrotron measurement (about 0.5\,wt.\%).

Temperature and field dependence of the dc magnetization was measured using the Quantum Design SQUID-VSM magnetometer (MPMS 3). Specific heat was measured in a Quantum Design Physical Properties Measurement System (QD-PPMS) using the relaxation method. The dc electrical resistivity was also measured in the PPMS using the standard four-probe technique. Electrical contacts were attached with the high-conducting silver paste. The two voltage leads were separated by 0.79\,mm, and the current was flowing through a cross-section of about 2.5 by 0.45\,mm$^2$. The powder was pressed into plate-like samples for the heat-capacity and resistivity measurements. 

High-field magnetization data up to 56\,T were collected in the Dresden High Magnetic Field Laboratory using a pulsed magnet. Experimental details of the measurement can be found elsewhere~\cite{tsirlin2009}. The collected high-field magnetization data were scaled using the magnetization data measured with the SQUID-VSM in static fields up to 7\,T.

Density-functional (DFT) band-structure calculations were performed within the \texttt{FPLO} code~\cite{fplo} using the experimental crystal structure determined at 20\,K and local density approximation (LDA) for the exchange-correlation potential~\cite{pw92}. {\cred The $8\times 8\times 8$ $k$-mesh was used for integration over the Brillouin zone. Additionally, we explored changes in the magnetic interactions in K$_2$IrCl$_6$ under the effect of strain that was modeled by changing the cubic lattice parameter and relaxing the Cl position until residual forces were below 0.001\,eV/\r A.}

Strong local correlations were included on the mean-field level via the DFT+$U$ procedure with the on-site Coulomb repulsion $U_d$ and Hund's coupling $J_d$ acting on the $5d$-states of Ir atoms, {\cred and the atomic-limit flavor of the double-counting correction.} Hopping parameters for the $t_{2g}$ states were extracted from the LDA band structure using Wannier functions implemented in \texttt{FPLO}.

\section{Results}
\subsection{Crystal structure}
\begin{figure}
\includegraphics[width=8.5cm]{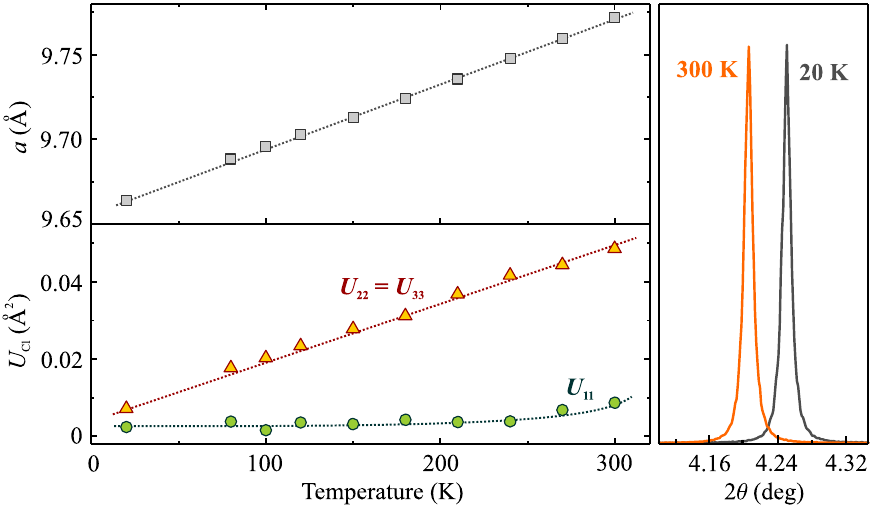}
\caption{\label{fig:xrd}
Left panel: temperature dependence of the cubic lattice parameter ($a$) and anisotropic atomic displacement parameters for Cl ($U_{\rm Cl}$). The lines are guide for the eye. As Cl occupies the $(x,0,0)$ position, the $U_{22}=U_{33}\gg U_{11}$ regime indicates an elongation of the displacement ellipsoid perpendicular to the Ir--Cl bond due to a soft rotary mode. Right panel: the 111 reflection measured by synchrotron XRD at 20 and 300\,K (ALBA, $\lambda=0.4129$\,\r A)
}
\end{figure}

Many of the K$_2$PtCl$_6$-type compounds undergo symmetry-lowering transitions upon cooling~\cite{roessler1977}. Therefore, we verified the cubic symmetry of K$_2$IrCl$_6$ using high-resolution synchrotron XRD and also refined the crystal structure, as no structural information was available in the literature. No deviations from the face-centered cubic symmetry are observed down to 20\,K (Fig.~\ref{fig:xrd}), and the temperature-independent peakwidth of $0.007^{\circ}$ for the 111 reflection suggests excellent crystallinity of the sample. Moreover, the absence of any thermodynamic anomalies below 20\,K and down to the magnetic ordering transition (see below) implies that K$_2$IrCl$_6$ should retain its cubic symmetry down to at least $T_N\simeq 3$\,K. 

\begin{table}
\caption{Crystallographic parameters and details of the structure refinement for K$_2$IrCl$_6$ at 300\,K and 20\,K. The atomic displacement parameters $U_{\rm iso}$ are given in \r A$^2$. The error bars are from the Rietveld refinement. All the crystallographic sites are fully occupied.
}\label{tab:structure}
\begin{ruledtabular}
\begin{tabular}{c c c}
$T(K)$ & 300 K & 20 K \\ 
\hline 
Space group & $Fm\bar 3m$ & $Fm\bar 3m$ \\ 
\hline 
$a=b=c$\,({\AA}) & 9.77050(3) & 9.66289(3) \\ 
\hline 
$\alpha=\beta=\gamma$ & 90$^{\circ}$ & 90$^{\circ}$ \\ 
\hline 
$V$({\AA}$^3$) & 932.718(5) & 902.238(4) \\ 
\hline 
$R_{I}/R_{P}$ & 0.0400/0.0902 & 0.0223/0.0928 \\ 
\hline 
Atomic parameters &  &  \\ 

 & $x/a=0$ & $x/a=0$ \\ 

Ir& $y/b=0$ & $y/b=0$ \\ 

 & $z/c=0$ & $z/c=0$ \\ 

 & $U_{\rm iso}=0.01647(9)$ & $U_{\rm iso}=0.00163(6)$ \\ 
\hline 
 & $x/c=\frac14$ & $x/a=\frac14$ \\ 

K & $y/b=\frac14$ & $y/b=\frac14$ \\ 

 & $z/c=\frac14$ & $z/c=\frac14$ \\ 

 & $U_{\rm iso}=0.0387(4)$  & $U_{\rm iso}=0.0061(2)$  \\ 
\hline 
 & $x/a=0.23708(10)$ & $x/a=0.24034(12)$ \\ 

Cl & $y/b=0$ & $y/b=0$ \\ 
 
 & $z/c=0$ & $z/c=0$ \\ 
 
 & $U_{\rm iso}=0.0357(3)$ & $U_{\rm iso}=0.0057(2)$ \\
\end{tabular}
\end{ruledtabular}
\end{table}

The lattice shrinks upon cooling, as seen from the refined lattice parameters, compare $a_{\rm 300 K}=9.77050(3)$\,{\AA} at 300\,K to $a_{\rm 20 K}=9.66289(3)$\,{\AA} at 20\,K. This corresponds to a 3\% volume reduction. Potential signatures of the symmetry lowering can be seen in the relatively high atomic displacement parameters of K and Cl at 300\,K (Table~\ref{tab:structure}). However, both displacements are significantly reduced upon cooling and drop well below 0.01\,\r A$^2$ at 20\,K (Fig.~\ref{fig:xrd}), suggesting the presence of a soft phonon mode but no static disorder. This is notably different from another Ir-based fcc antiferromagnet, the metrically cubic perovskite Ba$_2$CeIrO$_6$, where atomic displacements of oxygen remain well above 0.01\,\r A$^2$ even at 100\,K indicating static local disorder~\cite{revelli2019}. 

Thermal ellipsoid of Cl is stretched along the direction perpendicular to the Ir--Cl bond (Fig.~\ref{fig:xrd}). This would be typical for a rotary mode (cooperative rotations of the IrCl$_6$ octahedra), which is indeed common among the A$_2$IrX$_6$ antifluorite compounds~\cite{roessler1977,lynn1978,armstrong1980}. The reduction in the displacements upon cooling indicates the gradual suppression of such a mode in K$_2$IrCl$_6$, and underpins the absence of local distortions in this cubic compound at low temperatures.

The Ir--Cl distance changes from 2.3164(10)\,{\AA} at 300\,K to 2.3224(11)\,{\AA} at 20\,K. The Cl--Ir--Cl angles are fixed at $90^{\circ}$ by the cubic symmetry, resulting in the regular IrCl$_6$ octahedra.

\subsection{Resistivity}
\begin{figure}
	\centering
	\includegraphics[scale=0.5]{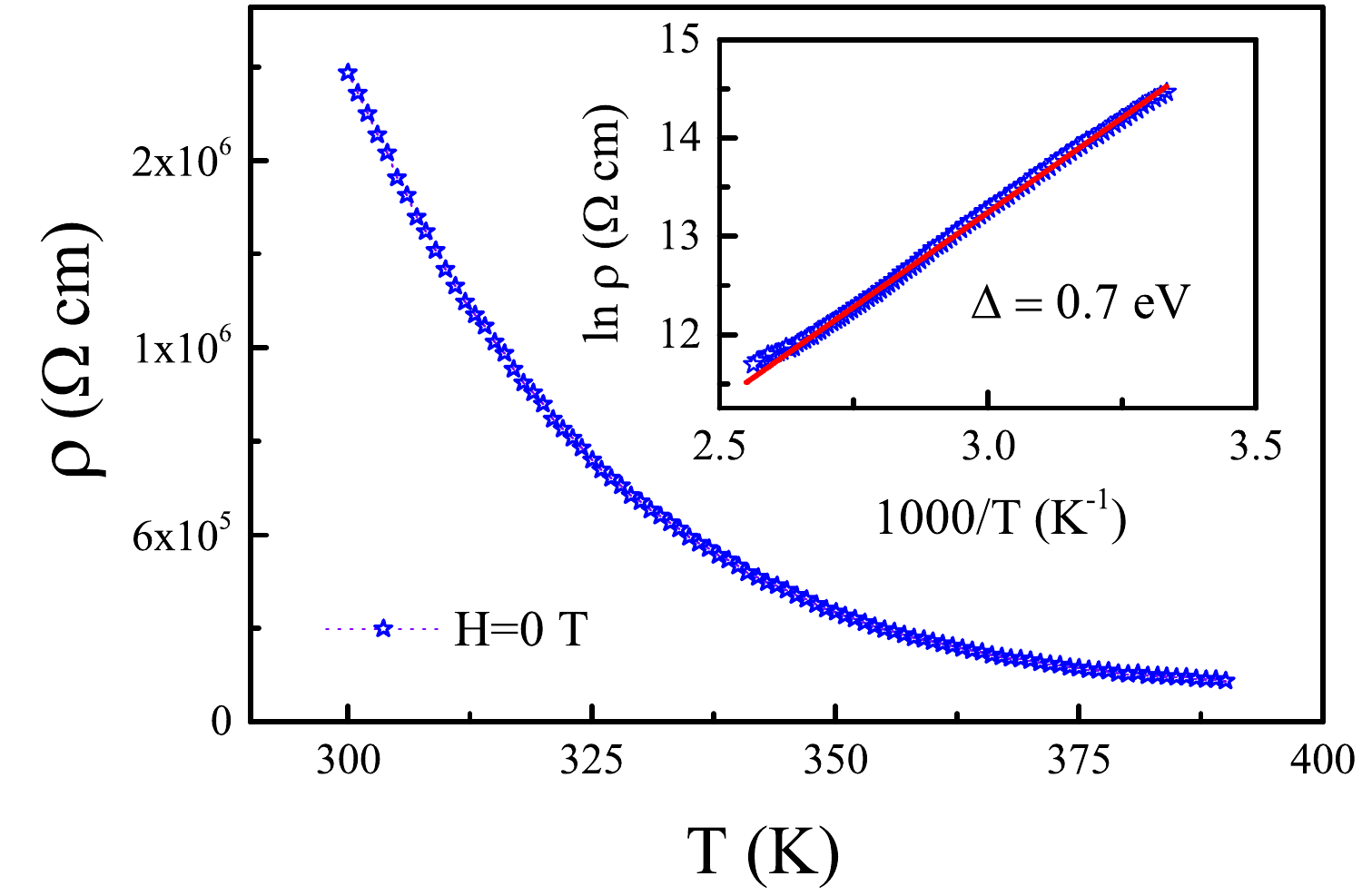}
	\caption{Zero-field electrical resistivity as a function of temperature for K$_2$IrCl$_6$. The inset shows the $\ln\rho$ vs $1/T$ plot, and the solid line is the linear fit.}\label{Fig.1}
\end{figure}

K$_2$IrCl$_6$ is a robust insulator. Its resistivity increases upon cooling from 390 K to 300\,K and exceeds 2\,M$\Omega$\,cm at room temperature. This confirms the anticipated Mott-insulating nature of the compound. The inset of Fig.~\ref{Fig.1} shows that the $T$ dependence of $\rho$ is well described by the activation behavior 
\begin{equation}
 \rho(T)=\exp\left(\frac{\Delta}{2k_B T}\right),
\end{equation}
where $\Delta$ is the activation energy and $k_B$ is the Boltzmann constant. From the linear fit to the $\ln\rho$ vs $1/T$ curve, we estimate $\Delta\sim 0.7$\,eV. The $\Delta$ value for K$_2$IrCl$_6$ is larger than that for the Ir$^{4+}$ oxides, such as Na$_2$IrO$_3$ ($\Delta\simeq 0.35$\,eV)~\cite{manni} and La$_2$MgIrO$_6$ ($\Delta\simeq 0.16$\,eV)~\cite{cao2013}. {\cred Although the exact $\Delta$ value for the polycrystalline sample may be affected by grain boundaries, the difference from the Ir$^{4+}$ oxides appears large enough to conclude that K$_2$IrCl$_6$ demonstrates higher ionicity. This is compatible with the results of our computational analysis presented in Sec.~\ref{sec:dft} below.}

\subsection{Magnetization}
\begin{figure}
	\centering
	\includegraphics[scale=0.49]{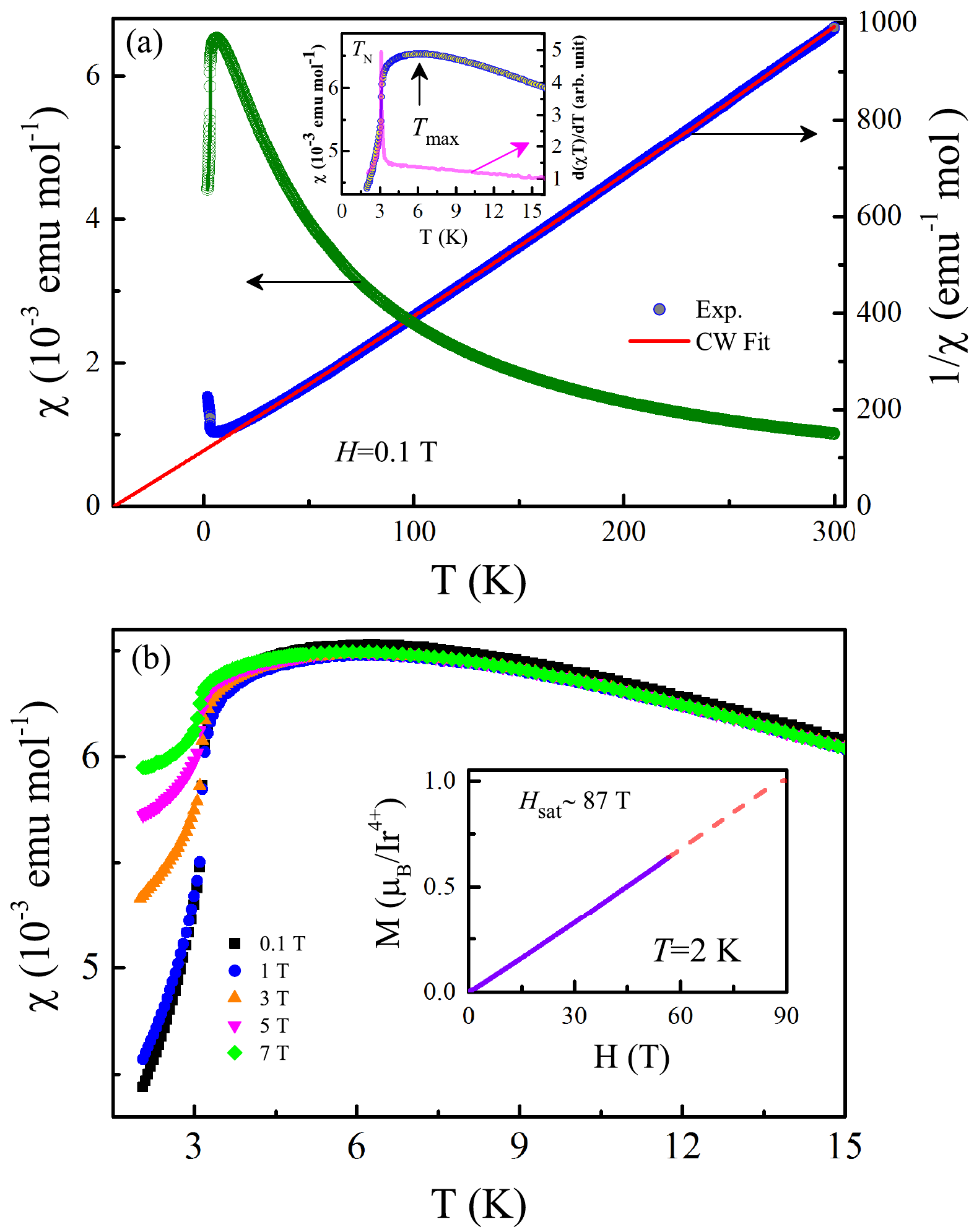}
	\caption{Temperature dependence of dc-magnetic susceptibility ($\chi=M/H$) and inverse magnetic susceptibility ($\chi^{-1}$) for K$_2$IrCl$_6$ measured under applied magnetic field $H=0.1$\,T. The solid line is the Curie-Weiss fit to the $\chi^{-1}(T)$ data at $100\leq T\leq 300$\,K. The inset shows the $\chi(T)$ and $d(\chi T)/dT$ plot at low temperatures. (b) $\chi$ vs $T$ plots measured under different applied magnetic fields. The inset shows the field dependence of dc-magnetization measured at 2\,K, and the dotted line is the linear extrapolation to the saturation magnetization $M_s=1$\,$\mu_B$/f.u. expected for $J=\frac12$ and $g=2$.}\label{Fig.2}
\end{figure}

The temperature-dependent dc-magnetic susceptibility $\chi=M/H$ measured under the $H=0.1$\,T applied magnetic field is shown in Fig.~\ref{Fig.2}a. Upon cooling, the susceptibility curve exhibits a broad maximum at $T_{\rm max}\sim$ 6.0\,K implying the onset of short-range spin-spin correlations. Below 6.0\,K, the susceptibility decreases down to 2\,K with the maximum in the Fisher's heat capacity $d(\chi T)/dT$ located around 3.1\,K, where a magnetic transition was reported in previous studies~\cite{cooke1959,minkiewicz1968}. The susceptibility data above 100\,K have been fitted using the following expression,
\begin{equation}
\chi=\chi_{0}+\frac{C}{T-\theta_{\rm CW}},\label{Eq.1}
\end{equation}
where $\chi_0$ is the temperature-independent contribution due to the core diamagnetism ($\chi_{\rm core}$) and Van Vleck paramagnetism ($\chi_{\rm VV}$). The second term represents the Curie-Weiss law with the Curie constant $C$ and Curie-Weiss temperature $\theta_{\rm CW}$. The Curie constant is given by $C=N_{\rm A}\mu_{\rm eff}^2/3k_{\rm B}$, where $N_{\rm A}$ is Avogadro's number, $\mu_{\rm eff}$ is the effective magnetic moment, and $k_{\rm B}$ is the Boltzmann constant.

The least-square fitting with Eq.~\eqref{Eq.1} above 100\,K returns $\chi_0=-8.53(9)\times 10^{-5}$\,emu/mol, $C=0.374(1)$\,emu\,K/mol, and $\theta_{\rm CW}=-42.6(1)$\,K. The negative value of $\theta_{\rm CW}$ implies predominant AFM exchange interaction between the Ir$^{4+}$ ions in K$_2$IrCl$_6$. The frustration parameter, $f=|\theta_{\rm CW}/T_{\rm N}|$, is estimated to be 13.7 and suggests the presence of strong magnetic frustration in K$_2$IrCl$_6$. 

The effective magnetic moment, $\mu_{\rm eff}=1.73$\,$\mu_{\rm B}$ is in excellent agreement with the expected value $\mu_{\rm eff}=g\sqrt{J(J+1)}\mu_{\rm B}$ using the Land\'e $g$-factor $g=2$ for the ideal $j_{\rm eff}=\frac12$ state expected for Ir$^{4+}$ in the cubic crystal field. The temperature-independent contribution $\chi_0$ includes the Van Vleck part $\chi_{\rm VV}$ and the core part $\chi_{\rm core}$=$-$2$\times$10$^{-4}$\,emu/mol~\cite{bain2008}. The Van Vleck susceptibility is then evaluated by subtracting $\chi_{\rm core}$ from $\chi_0$, resulting in $\chi_{\rm VV}$= 1.1$\times$10$^{-4}$\,emu/mol very similar to $1.05\times 10^{-4}$\,emu/mol reported for Na$_2$IrO$_3$~\cite{mehlawat2017}.

Figure~\ref{Fig.2}(b) shows temperature-dependent susceptibility measured under different magnetic fields. Below $T_{\rm max}$, the susceptibility increases with increasing the field, and the transition at $T_N$ is gradually smeared out, but can be still observed in $d(\chi T)/dT$. The ordering temperature $T_N$ changes insignificantly. For example, the peak in $d(\chi T)/dT$ shifts from 3.14\,K at 0.01\,T to 3.07\,K at 7\,T~\cite{supplement}. 

Linear field dependence of the magnetization measured at 1.6\,K (Fig.~\ref{Fig.2}(b)) persists up to 56\,T, the highest field of our experiment. The scaling against the low-field data measured in static fields suggests that at 56\,T the magnetization reaches 63.8\% of the expected saturation value of $M_s=gS\mu_{\rm B}=1.0$\,$\mu_{\rm B}$/Ir$^{4+}$. Linear extrapolation to $M_s$ yields the saturation field of $H_{\rm sat}\sim 87$\,T. 

\subsection{Heat capacity}
\begin{figure}
	\centering
	\includegraphics[scale=0.5]{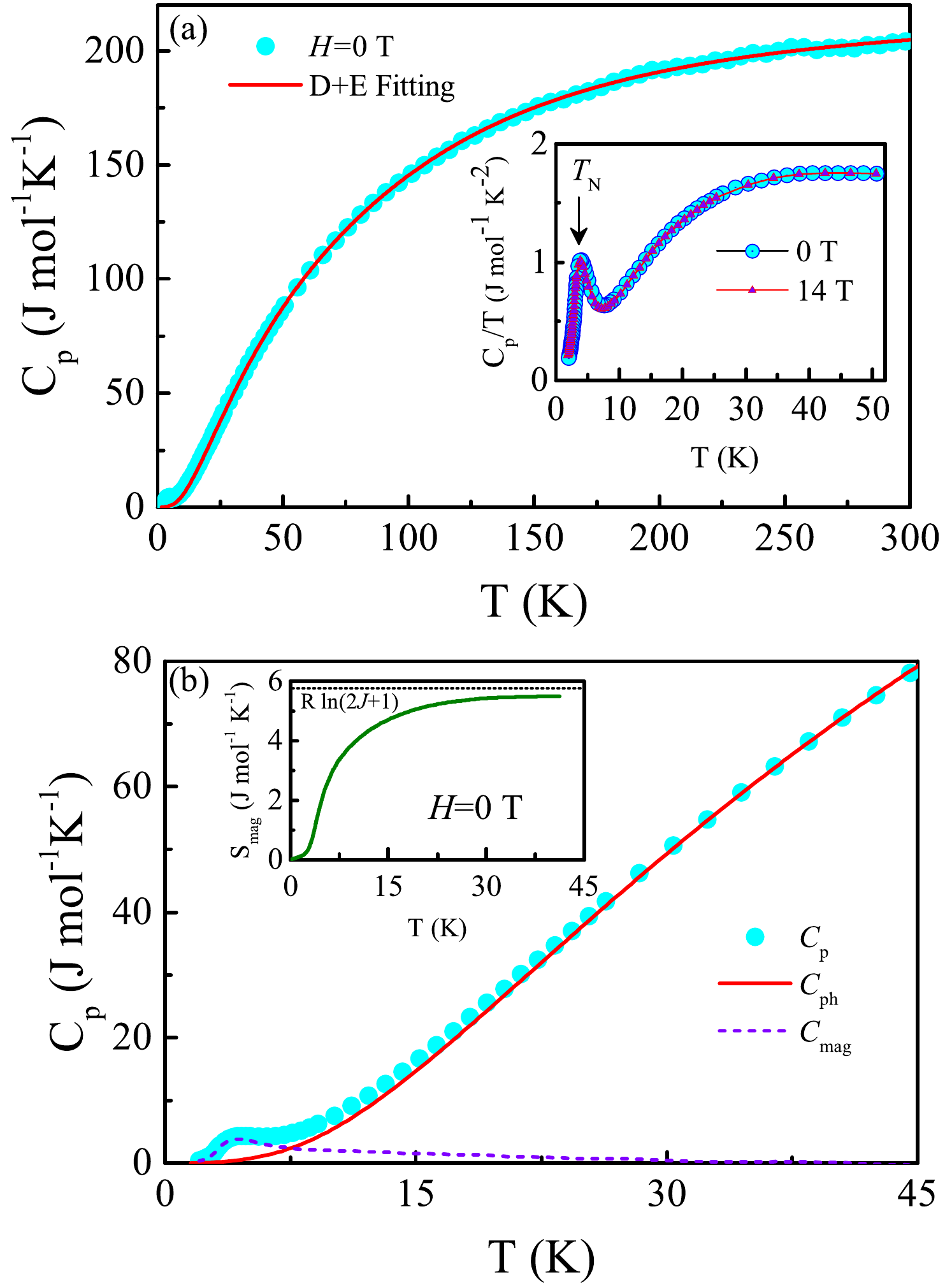}
	\caption{(a) Temperature dependence of the zero-field specific heat ($C_p$) for K$_2$IrCl$_6$ measured from 300\,K to 2\,K. Open circles are the raw data, and solid red line is the least-square fit to the lattice specific heat following Eq.~\eqref{Eq.2}. The inset shows the $C_p/T$ as a function of $T$ for the $H=0$ and 14\,T applied magnetic fields, exhibiting clear anomaly near the transition temperature $T_N$. (b) The $T$-dependence of the raw specific heat $C_p$ (open circles), the lattice contribution $C_{\rm ph}$ (solid line), and the magnetic contribution $C_{\rm mag}$ (dashed line) in the low-temperature region. The inset shows the magnetic entropy $S_{\rm mag}$ as a function of $T$ for $H=0$\,T. The dotted line refers to the theoretically predicted magnetic entropy $S_{\rm mag}=R\ln(2J+1)$.}\label{Fig.3}
\end{figure}

Fig.~\ref{Fig.3}(a) shows the heat capacity of K$_2$IrCl$_6$. A broad hump around 5\,K is due to the magnetic contribution, which remains unchanged even in the applied field of 14\,T. This field only weakly polarizes the system to produce about 15\,\% of the maximum magnetization $M_s$, thus having no significant effect on the magnetic contribution to the specific heat. Interestingly, neither our magnetization data nor the specific heat reveal any signatures of the field-induced transition reported in Ref.~\onlinecite{meschke2001} at about 5\,T. However, these signatures may be weak and not easily resolvable in a polycrystalline sample.

To determine the magnetic contribution to the specific heat, the phonon part $C_{\rm ph}(T)$ was estimated by fitting experimental heat capacity above 35\,K with an empirical model that involves a superposition of one Debye-type and three Einstein-type terms as follows~\cite{koteswararao2014}: 
\begin{equation}
C_{\rm ph}(T)=f_{D}C_{D}(\Theta_{D},T)+\sum_{i=1}^{3}  g_{i}C_{E,i}(\Theta_{E,i},T).\label{Eq.2}
\end{equation}
The Debye term  $C_{D}(\Theta_{D},T)$ is given by  
\begin{equation}
C_{D}(\Theta_{D},T)= 9R\left (\frac{T}{\Theta_{D}}\right)^3\int_{0}^{\Theta_{D}/T}\frac{x^4 e^x}{(e^x -1)^2}dx, 
\end{equation}
and the Einstein term $C_{E,i}(\Theta_{E,i},T)$ is given by
\begin{equation}
C_{E,i}(\Theta_{E,i},T)= 3R\left(\frac{\Theta_{E, i}}{T}\right)^2\frac{\exp(\Theta_{E,i/T})}{[\exp(\Theta_{E,i/T})-1]^2}, 
\end{equation}
where $R$ is the universal gas constant, $k_{B}$ is the Boltzmann constant, $\Theta_{D}$ and $\Theta_{E}$ are the Debye and Einstein temperatures, respectively. In this combined Debye-Einstein ($D+E$) model, the total number of vibration modes $n=9$ is the total number of atoms in the formula unit. 

A stable least-square fit of Eq.~\eqref{Eq.2} to the $C_{p}(T)$ data above $T=35$\,K has been obtained by a combination of one Debye term and three Einstein terms with 24\% of the total modes contributed by the Debye term [Fig.~\ref{Fig.3}(a)]. During the fitting procedure, the Debye temperature $\Theta_{D}$ was kept fixed to that reported for the iso-structural non-magnetic K$_2$PtCl$_6$ compound, $\Theta_{D}$=91.1(5)\,K~\cite{moses1979}. The fit yields three Einstein temperatures corresponding to the three Einstein terms of the fitted model as $\Theta_{E}$= 459(7), 242(4), and 136(2)\,K. These Einstein modes can be compared to the relatively flat phonon modes in the isostructural K$_2$OsCl$_6$ in the energy range between 150 and 250\,K~\cite{sutton1983}, whereas the relatively low (effective) Debye temperature may be caused by the aforementioned soft rotary mode~\cite{mintz1979,lynn1978}. This interpretation compares favorably to the results of the structure analysis that reveal strong temperature dependence of the atomic displacements parameters for K and Cl (Table~\ref{tab:structure}).

By subtracting $C_{\rm ph}(T)$ from the $C_{p}(T)$ data, we obtain the temperature-dependent magnetic contribution $C_{\rm mag}$ as shown in Fig. \ref{Fig.3}(b). It reveals a broad maximum around 5\,K comparable to $T_{\max}\simeq 6$\,K in the magnetic susceptibility, and gradually decreases toward higher temperatures. From $C_{\rm mag}(T)$ data, the total magnetic entropy is estimated as follows 
\begin{equation}
S_{\rm mag}(T)=\int_{0}^{T}\frac{C_{\rm mag}}{T'}dT',
\end{equation}   
with the high-temperature limit of 5.50 J\,mol$^{-1}$\,K$^{-1}$. The proximity to the theoretical value of $S_{\rm mag}=R\ln(2J+1)=5.76$\,J\,mol$^{-1}$\,K$^{-1}$ validates our analysis. 

The maximum value of $C_{\rm mag}$ is about $0.45R$ and comparable to that in non-frustrated square-lattice antiferromagnets~\cite{bernu2001}. This value gauges the effect of quantum fluctuations, because both low-dimensionality and magnetic frustration tend to impede short-range order and reduce the maximum in $C_{\rm mag}$. Our experimental value for K$_2$IrCl$_6$ suggests that the geometrical and exchange frustration of the three-dimensional fcc lattice in K$_2$IrCl$_6$ cause quantum effects that are as strong as in square-lattice antiferromagnets, but weaker than in frustrated two-dimensional systems, such as triangular antiferromagnets with the maximum value of $C_{\rm mag}$ of $0.22R$ only~\cite{bernu2001}.

In contrast to previous studies~\cite{bailey1959,moses1979}, we do not observe a sharp transition anomaly at $T_N$ and rather detect a broad maximum of $C_{\rm mag}$ around this temperature. The sharp anomaly in $d(\chi T)/dT$ at 3.1\,K measured on the same sample shows that the absent specific-heat anomaly is not a drawback of sample quality. Indeed, we repeated specific-heat measurement on a different sample, but the anomaly remained broad.

\subsection{Electronic structure}
\label{sec:dft}
\begin{figure}
\includegraphics[width=\columnwidth]{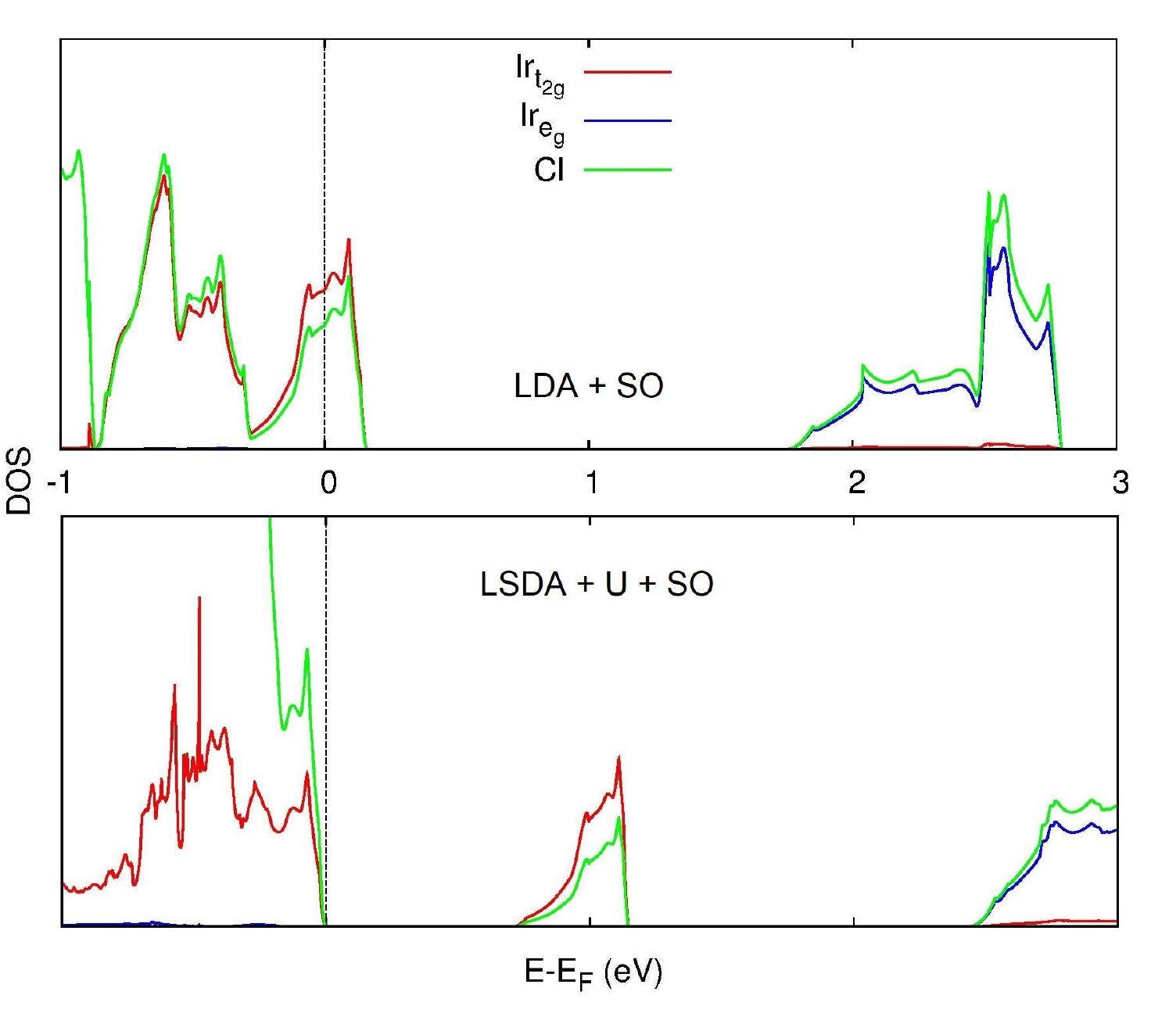}
\caption{\label{fig:elstr}
Atomic- and orbital-resolved DOS from the LDA+SO (top) and ferromagnetic LSDA+$U$+SO (bottom) calculations with $U_d=2.2$\,eV {\cred and $J_d=0.3$\,eV.} The Fermi level is at zero energy.
}
\end{figure}

We now proceed to the computational analysis. The uncorrelated (LDA+SO) electronic structure of K$_{2}$IrCl$_{6}$ shown in Fig.~\ref{fig:elstr} reveals a combination of Ir $t_{2g}$ and Cl $3p$ states at the Fermi level. The $t_{2g}$ bands between $-1$ and $0.5$\,eV develop the characteristic two-peak structure that corresponds to the splitting of $t_{2g}$ states into the $j=\frac{1}{2}$ and $j=\frac{3}{2}$ levels.

Correlation effects in the Ir $5d$ shell are taken into account within LSDA+$U$+SO. The on-site Coulomb repulsion $U_d=1.7$\,eV and Hund's exchange $J_d=0.3$\,eV are commonly used for the Ir$^{4+}$ oxides~\cite{winter2016,winter2017}. This set of parameters leads to a band gap of 0.5\,eV, which is even lower than the activation energy of $\Delta\simeq 0.7$\,eV for the electrical transport. The experimental value of $\Delta$ is well reproduced with $U_d=2.2$\,eV that we choose as the optimal value for K$_2$IrCl$_6$. In Sec.~\ref{sec:model}, we show that the very same value of $U_{\eff}=U_d=2.2$\,eV leads to a good agreement with the experimental Curie-Weiss temperature and saturation field and thus can be used for the evaluation of magnetic parameters. 

The increased on-site Coulomb repulsion reflects the weaker screening by ligands and the higher ionicity of K$_2$IrCl$_6$ in agreement with the larger $\Delta$, which is unusually high for an Ir$^{4+}$ compound. Interestingly, this increased ionicity is not immediately visible in the atomic-resolved LDA+SO density of states (DOS), where Cl $3p$ orbitals contribute about 44\,\% of the total DOS at the Fermi level, an even a larger contribution than 34\,\% of O $2p$ states in Na$_2$IrO$_3$. 

\begin{figure}
\includegraphics[width=\columnwidth]{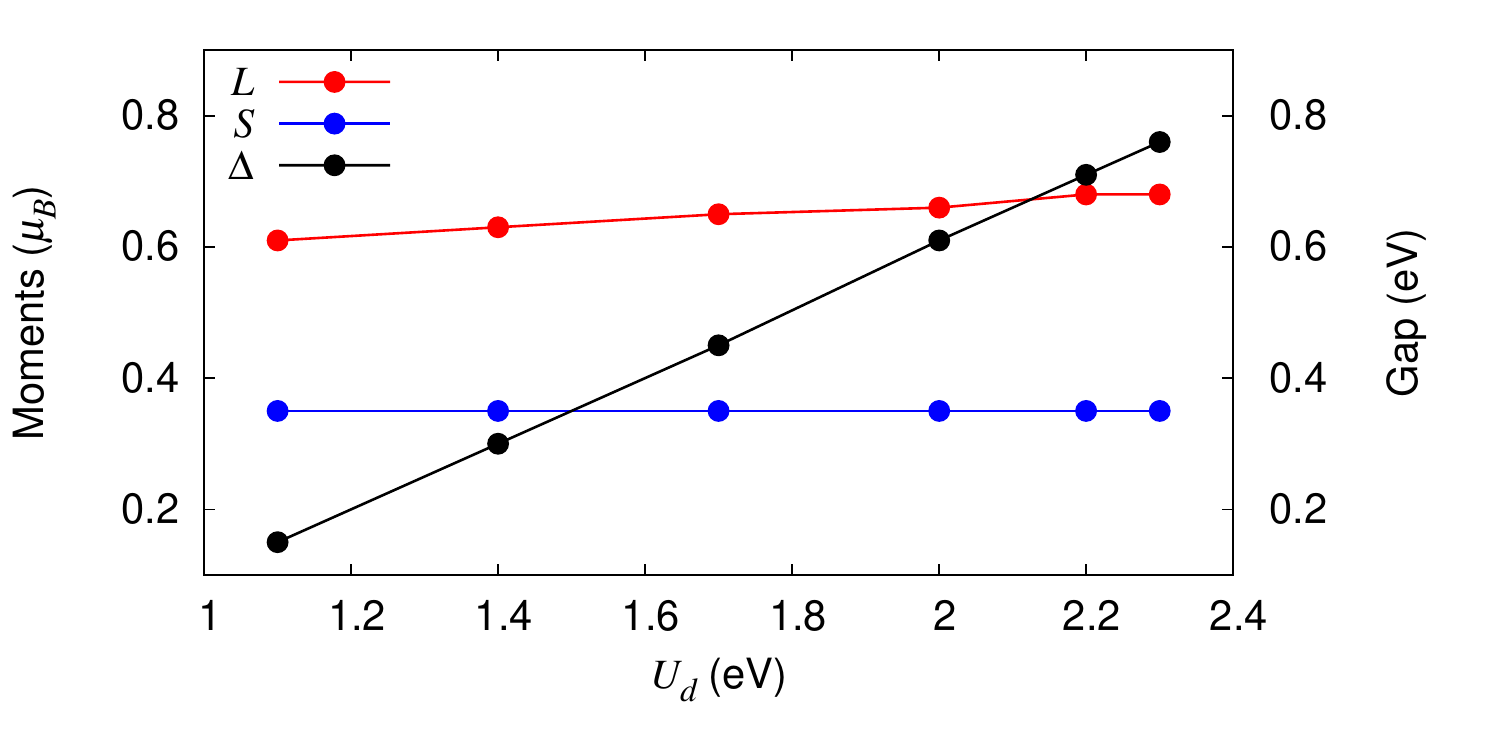}
\caption{\label{fig:uparam}
Band gap $\Delta$ (left) and spin ($S$) as well as orbital ($L$) moments of Ir (right) as a function of $U_d$ {\cred calculated with the constant $J_d=0.3$\,eV}. The spin and orbital moments are given in $\mu_B$ and obtained in a ferromagnetic configuration by summing the contributions from all atoms. }
\end{figure}

The spin and orbital moments obtained in LSDA+$U$+SO are consistent with the anticipated $j_{\rm eff}=\frac12$ state of Ir$^{4+}$. We find the spin moment of 0.33\,$\mu_B$ and the orbital moment of $0.6-0.7$\,$\mu_B$ only weakly dependent on the $U_d$ value (Fig.~\ref{fig:uparam}, right).

\subsection{Microscopic magnetic model}
\label{sec:model}

Exchange couplings in Ir$^{4+}$ compounds are generally anisotropic. The spin Hamiltonian can be written as
\begin{equation}
H=\sum_{\langle ij\rangle} \mathbf S_i\,\mathbb J_{ij}\,\mathbf S_j,
\end{equation}
with the sum taken over all pairs of atoms. The $\mathbb J_{ij}$'s are exchange tensors of the form:
\begin{equation}
\mathbb J_{ij}=
\begin{pmatrix}
J+\Gamma_{xx} & D_{z}+\Gamma_{xy}  & -D_{y}+\Gamma_{xz} \\
  -D_{z}+\Gamma_{yx} & J+\Gamma_{yy} & D_{x}+\Gamma_{yz} \\
  D_{y}+\Gamma_{zx} & -D_{x}+\Gamma_{zy} & J+\Gamma_{zz}
\end{pmatrix}
,
\notag\end{equation}
where $J$ is the isotropic (Heisenberg) coupling, $D$ is the Dzyaloshinskii-Moriya interaction vector, and $\Gamma$ is the second-rank traceless tensor that describes the symmetric portion of the anisotropic exchange.

The cubic symmetry of the structure leads to the following form of the exchange tensor for three groups of nearest-neighbor interactions,
\begin{equation}
\mathbb J_{xy}=
\begin{pmatrix}
J & \pm \Gamma  &0 \\
  \pm \Gamma & J & 0\\
  0 & 0 & J+K
\end{pmatrix}
\notag
\end{equation}
\begin{equation}
\mathbb J_{xz}=
\begin{pmatrix}
J & 0  & \pm \Gamma \\
  0 & J+K & 0\\
  \pm \Gamma & 0 & J
\end{pmatrix} 
\notag\end{equation}
\begin{equation}
\mathbb J_{yz}=
\begin{pmatrix}
J+K & 0  &0 \\
  0 & J & \pm \Gamma\\
  0 & \pm \Gamma& J
\end{pmatrix}
. \notag
\end{equation}
Here, $\mathbb J_{xy}$, $\mathbb J_{xz}$, and $\mathbb J_{yz}$ stand for exchange tensors of the bonds on the respective faces of the cubic unit cell (Fig.~\ref{fig:structure}, right). All components of the $\Gamma$-tensor are reduced to only two parameters, the diagonal (Kitaev) exchange $K$ and the off-diagonal anisotropy $\Gamma$, whereas Dzyloshinskii-Moriya interaction vanishes, owing to the inversion symmetry of the nearest-neighbor Ir--Ir exchange bonds.

{\cred To estimate the $J$, $K$, and $\Gamma$ parameters, we use the perturbation-theory approach detailed in Refs.~\onlinecite{rau2014,winter2016}. LDA hoppings within the $t_{2g}$ manifold form the hopping matrix
\begin{equation}
\mathbb T=
\begin{pmatrix}
 t_1 & t_2 & t_4 \\
 t_2 & t_1 & t_4 \\
 t_4 & t_4 & t_3 \\
\end{pmatrix}
\notag
\end{equation}
written in the $d_{yz}-d_{xz}-d_{xy}$ basis, respectively. Magnetic interaction parameters are obtained as~\cite{winter2016}
\begin{gather}
J=\frac{4\mathbb A}{9}\left ( 2t_1+t_3 \right )^2-\frac{8\mathbb B}{9}\left \{ 9t_4^2+ 2(t_1-t_3)^2 \right \} \\
K=\frac{8\mathbb B}{3}\left \{(t_1-t_3)^2+3t_4^2-3t_2^2 \right \}
\label{eq:kitaev} \\
\Gamma=\frac{8\mathbb B}{3}\left \{2t_2(t_1-t_3)+3t_4^2 \right \}
\end{gather}
using the constants
\begin{gather*}
\mathbb A = -\frac{1}{3}\left \{ \frac{J_H+3(U_{\rm eff}+3\lambda)}{6J_H^2-U_{\rm eff}(U_{\rm eff}+3\lambda)+J_H(U_{\rm eff}+4\lambda)} \right \} \\[5pt]
\mathbb B = \frac{4}{3}\left \{ \frac{(3J_H-U_{\rm eff}-3\lambda)}{(6J_H-2U_{\rm eff}-3\lambda)}\eta \right \} \\[5pt]
\eta = \frac{J_H}{6J_H^2-J_H(8U_{\rm eff}+17\lambda)+(2U_{\rm eff}+3\lambda)(U_{\rm eff}+3\lambda)}.
\end{gather*}
}

Using the spin-orbit coupling $\lambda=0.4$\,eV, Hund's coupling $J_H=J_d=0.3$\,eV, as well as the effective Coulomb repulsion $U_{\eff}=2.2$\,eV determined in the previous section, we arrive at $J=13$\,K, $K=5$\,K, and $\Gamma=1$\,K. The calculated Curie-Weiss temperature $\Theta=-(3J+K)=-44$\,K is in good agreement with the experimental value of $\Theta=-42.6(1)$\,K. Moreover, our \textit{ab initio} $J$ compares favorably to $J\simeq 11.5$\,K extracted from electron spin resonance experiments on the magnetically diluted samples~\cite{griffiths1959}. The same experiments provide an estimate for the exchange anisotropy $J^z-(J^x+J^y)/2\simeq 2$\,K~\cite{griffiths1959}, which is comparable to our $K$, although one should keep in mind that Ref.~\onlinecite{griffiths1959} assumed the conventional orthorhombic exchange anisotropy instead of the actual Kitaev one.

\begin{figure}
\includegraphics[width=\columnwidth]{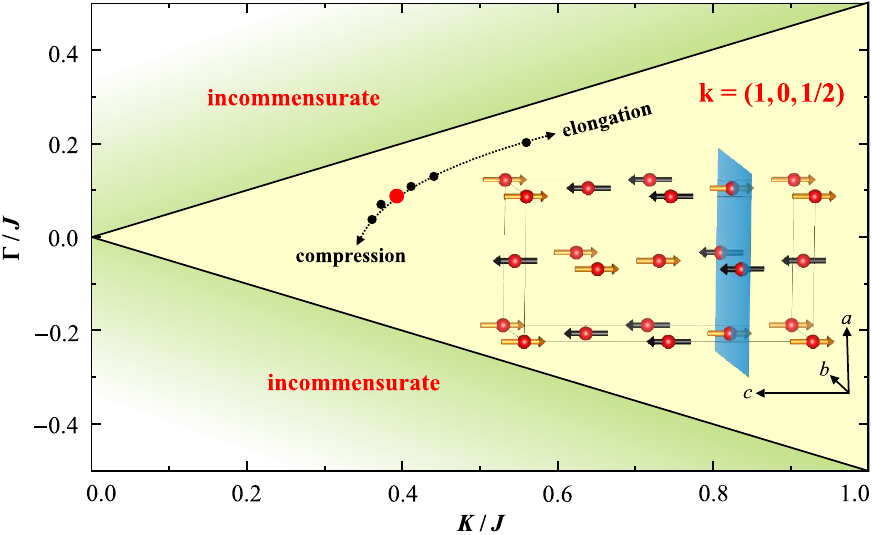}
\caption{\label{fig:phased}
Classical phase diagram of the $J-K-\Gamma$ model and the location of K$_2$IrCl$_6$ therein (larger red point). Smaller black points show the evolution upon compressing the structure to $a=9.17$\,\r A or expanding it to $a=11.37$\,\r A~\cite{supplement}. The experimental $\mathbf k=(1,0,\frac12)$ order is shown, with one of the antiferromagnetic $ab$ planes highlighted. 
}
\end{figure}

The ratios of $K/J\simeq 0.38$ and $\Gamma/J\simeq 0.08$ would place K$_2$IrCl$_6$ into the region of collinear antiferromagnetic order with the propagation vector $\mathbf k=(1,0,0)$ in the classical phase diagram for the $J-K-\Gamma$ model on the fcc lattice~\cite{cook2015}. However, experimental neutron study~\cite{hutchings1967} revealed a different flavor of collinear antiferromagnetic order described by $\mathbf k=(1,0,\frac12)$. To resolve this discrepancy, we re-constructed the phase diagram (Fig.~\ref{fig:phased}) using the Luttinger-Tisza method, and recognized that these two states remain degenerate unless a second-neighbor interaction $J_2$ is included~\cite{haar1962,tahir1966}. Whereas a ferromagnetic $J_2$ would stabilize the $\mathbf k=(1,0,0)$ or type-I order, an antiferromagnetic $J_2$ leads to the $\mathbf k=(1,0,\frac12)$ or type-IIIA order~\footnote{For both types of order, cubic symmetry allows different choices of the propagation vector. Here, we assume the doubled periodicity along $c$ and, therefore, write $k_z=\frac12$, but for example $\mathbf k=(1,\frac12,0)$ of Ref.~\onlinecite{revelli2019} is equivalent to our $\mathbf k=(1,0,\frac12)$ with the $b$ and $c$ directions swapped.}. In our case, we find a weakly antiferromagnetic $J_2\simeq 0.2$\,K that would lead to the $\mathbf k=(1,0,\frac12)$ order observed experimentally. On the other hand, the $\mathbf k=(1,0,0)$ order has been experimentally observed in Ba$_2$CeIrO$_6$~\cite{aczel2019}, where it may be triggered by a weakly ferromagnetic $J_2$ or by local deviations from the cubic symmetry. 

The second-neighbor interaction $J_2$ remains very weak, $J_2/J\simeq 0.015$, as expected from the large Ir--Ir distance of nearly 10\,\r A. This DFT estimate is compatible with the remarkably low N\'eel temperature $T_N/J\simeq 0.24$ that both spin-wave~\cite{lines1963} and Green-function~\cite{lines1964} calculations predict in the region of $J_2/J_1<0.05$ only.

From the energy difference between the $\mathbf k=(1,0,\frac12)$ state and the fully polarized ferromagnetic state we estimate the saturation field of $H_s=(4J+2K)k_B/(g\mu_Bj_{\rm eff})\simeq 92$\,T in good agreement with our extrapolated value of 87\,T (Fig.~\ref{Fig.2}b). We further explored the stability of the $\mathbf k=(1,0,\frac12)$ order and performed \textit{ab initio} calculations for different values of the cubic lattice parameter of K$_2$IrCl$_6$~\cite{supplement}. Whereas compression leads to only minor changes in the exchange couplings and shifts the system deeper into the region of collinear order, an expansion of the structure would increase $K/J$. The classical phase boundary is not crossed, though.

The antiferromagnetic Kitaev exchange is somewhat uncommon, as honeycomb iridates and their analogs all show $K<0$~\cite{winter2017}. The origin of $K>0$ can be understood from Eq.~\eqref{eq:kitaev}. In K$_2$IrCl$_6$, $t_1=5.4$\,meV, $t_2=9.6$\,meV, $t_3=-82.9$\,meV, and $t_4=0$. Therefore, $K$ is mainly due to $t_3$ and antiferromagnetic, in contrast to the honeycomb iridates where large $t_2$ leads to $K<0$. As $t_3$ contributes to $J$ as well, the antiferromagnetic $K$ is necessarily supplemented by an even larger antiferromagnetic $J$ that occurs in K$_2$IrCl$_6$ indeed. Microscopically, the large-$t_2$ regime corresponds to the $90^{\circ}$ Ir--O--Ir superexchange between the edge-sharing IrO$_6$ octahedra, whereas in K$_2$IrCl$_6$ a longer Ir--Cl$\ldots$Cl--Ir superexchange pathway between the disconnected IrCl$_6$ octahedra leads to the large-$t_3$ regime caused by the Cl-mediated $d_{xy}-d_{xy}$ hopping. This microscopic mechanism is remarkably similar to the long-range superexchange in V$^{4+}$ compounds, where only the $d_{xy}$ orbital is magnetic~\cite{tsirlin2011}.

\section{Discussion and Summary}
Using high-resolution synchrotron x-ray diffraction, we verified the cubic symmetry of K$_2$IrCl$_6$ and the regular octahedral environment of Ir$^{4+}$. From crystallographic point of view, neither global nor local symmetry lowering would be expected in this compound. This is different from Ba$_2$CeIrO$_6$ that likely features local distortions revealed by the abnormally high displacement parameters of oxygen atoms~\cite{revelli2019} and from other Ir-based double perovskites, where even the symmetry of the average structure is lower than cubic~\cite{aczel2019}. While spectroscopy experiments would be needed to demonstrate the absence of the $t_{2g}$ crystal-field splitting and to ultimately confirm the $j_{\rm eff}=\frac12$ state of Ir$^{4+}$, we note that the results of such experiments may be temperature-dependent. Our data reveal dynamic distortions of the structure driven by the soft rotary mode. At elevated temperatures, this mode renders the system locally and instantaneously non-cubic. It may even be possible to observe not only the $j_{\rm eff}=\frac12$ state at low temperatures but also the gradual departure from it upon heating.

Our computational analysis backed by the results of thermodynamic and transport measurements reveals the increased on-site Coulomb repulsion compared to Ir$^{4+}$ oxides. The magnetic model comprises a sizable antiferromagnetic Kitaev exchange, albeit superimposed on an even larger Heisenberg term, which is unavoidable in this setting, because the leading hopping process is qualitatively different from that in honeycomb iridates with $K<0$ and small $J$.

The large frustration ratio of $\theta/T_N=13.7$ indicates a strongly impeded magnetic order, although its full suppression with the formation of a spin liquid appears impossible in the parameter range of K$_2$IrCl$_6$. Neither  compressive nor tensile strain moves the system sufficiently close to a classical phase boundary, where long-range order can be destabilized. That said, K$_2$IrCl$_6$ with its robust magnetic order appears to be a good model fcc antiferromagnet. An excellent match between the \textit{ab initio} results and experiment paves the way to studying excitations of this frustrated antiferromagnet and other properties that can be influenced by the frustration. One of them is the possible structural component in the magnetic ordering transition at $T_N$, or even the occurrence of two consecutive transitions at about 2.8 and 3.1\,K~\cite{moses1979}. Their detailed nature lies beyond the scope of our present study and requires dedicated experiments such as single-crystal neutron diffraction. Here, we only note that the absence of a clear transition anomaly in the specific heat of our polycrystalline samples serves as an additional evidence for the first-order nature of the transition(s) and, thus, for the presence of a structural component. This is not unexpected given the abundance of magnetic frustration and availability of soft phonon modes. Further work in this direction would be interesting. 

In summary, we confirmed the cubic symmetry of K$_2$IrCl$_6$ and detected a soft rotary mode that is gradually suppressed upon cooling. This compound is a geometrically frustrated fcc antiferromagnet with the nearest-neighbor Heisenberg exchange $J\simeq 13$\,K and Kitaev exchange $K\simeq 5$\,K augmented by a weak next-nearest-neighbor coupling $J_2\simeq 0.2$\,K that stabilizes the $\mathbf k=(1,0,\frac12)$ order. The activation energy of charge transport $\Delta\simeq 0.7$\,eV and the effective Coulomb repulsion $U_{\rm eff}=2.2$\,eV are higher than in Ir$^{4+}$ oxides, suggesting an increased ionicity of the chloride. The leading Ir--Ir hopping differs from that in the honeycomb iridates and triggers the antiferromagnetic Kitaev term accompanied by an even stronger Heisenberg one.

\acknowledgments
We acknowledge the provision of synchrotron beamtime by the ALBA and ESRF and thank Francois Fauth, Alexander Missyul, Oriol Vallcorba, Wilson Mogodi, and Mauro Coduri for their experimental assistance. We are also grateful to Anton Jesche for his help with the magnetization measurements and general advice, and to Philipp Gegenwart for useful comments. AT thanks Anna Efimenko, Liviu Hozoi, Vladimir Hutanu, Jeffrey Lynn, Sergey Zvyagin, and Adam Aczel for various communications on K$_2$IrCl$_6$. The support of the HLD at HZDR, member of the European Magnetic Field Laboratory (EMFL), is acknowledged. The work in Augsburg was supported by the Federal Ministry for Education and Research through the Sofja Kovalevskaya Award of Alexander von Humboldt Foundation. This work was funded by the
Deutsche Forschungsgemeinschaft (DFG, German Research Foundation) under Projektnummer 107745057 -- TRR 80 (Augsburg) and SFB1143 (Dresden-Rossendorf). The work of VGM was supported by the Russian Science Foundation Grant No. 18-12-00185. 

%

\clearpage\newpage

\renewcommand{\thefigure}{S\arabic{figure}}
\renewcommand{\thetable}{S\arabic{table}}
\setcounter{figure}{0}
\setcounter{table}{0}

\begin{figure*}
\includegraphics[width=12cm]{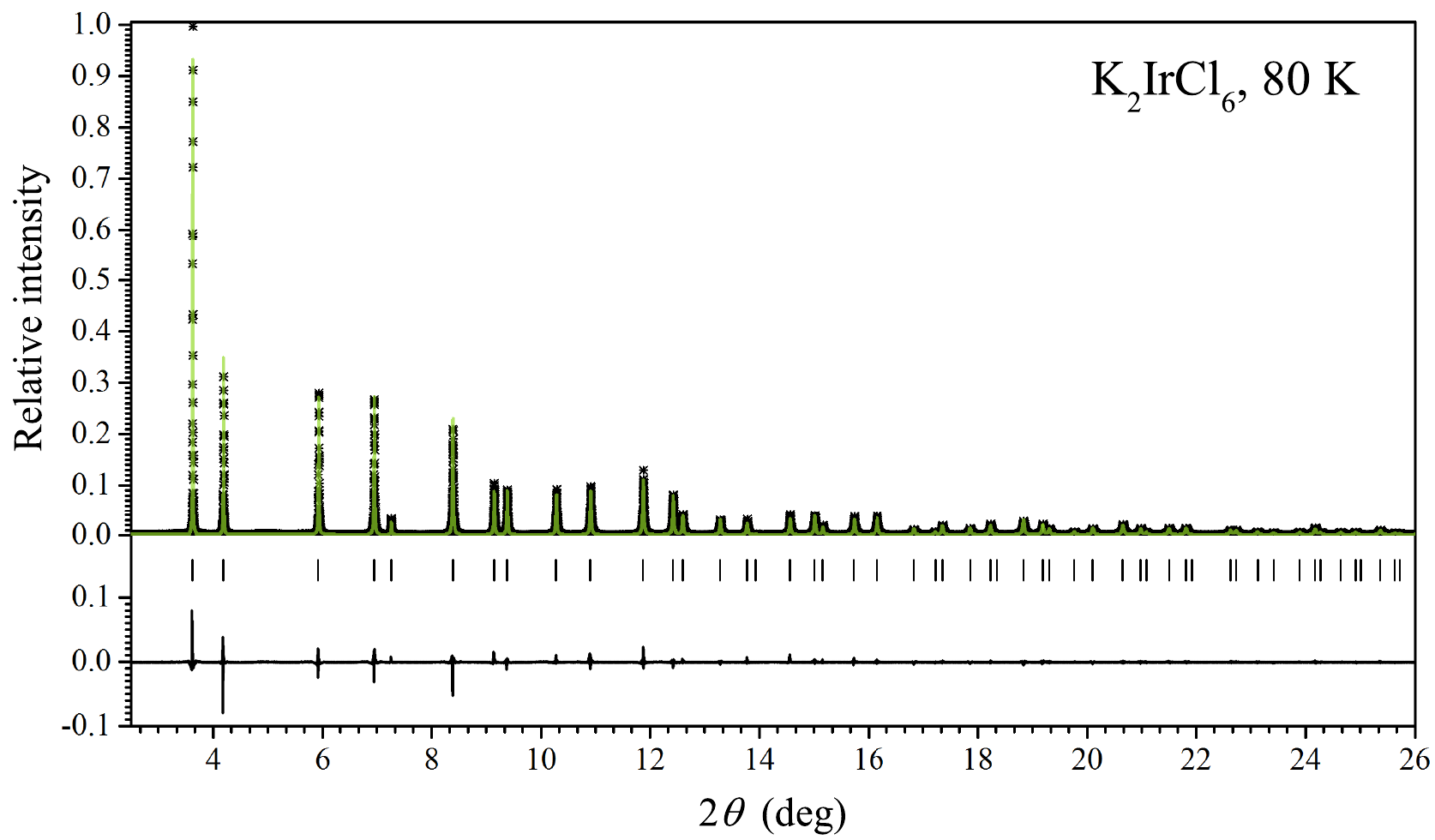}
\caption{
Rietveld refinement of the K$_2$IrCl$_6$ structure at 80\,K (ESRF, $\lambda=0.35456$\,\r A). Ticks show reflection positions expected for the cubic face-centered unit cell, whereas the line in the bottom of the figure is the difference pattern.
}
\end{figure*}

\begin{figure*}
\includegraphics[width=8.5cm]{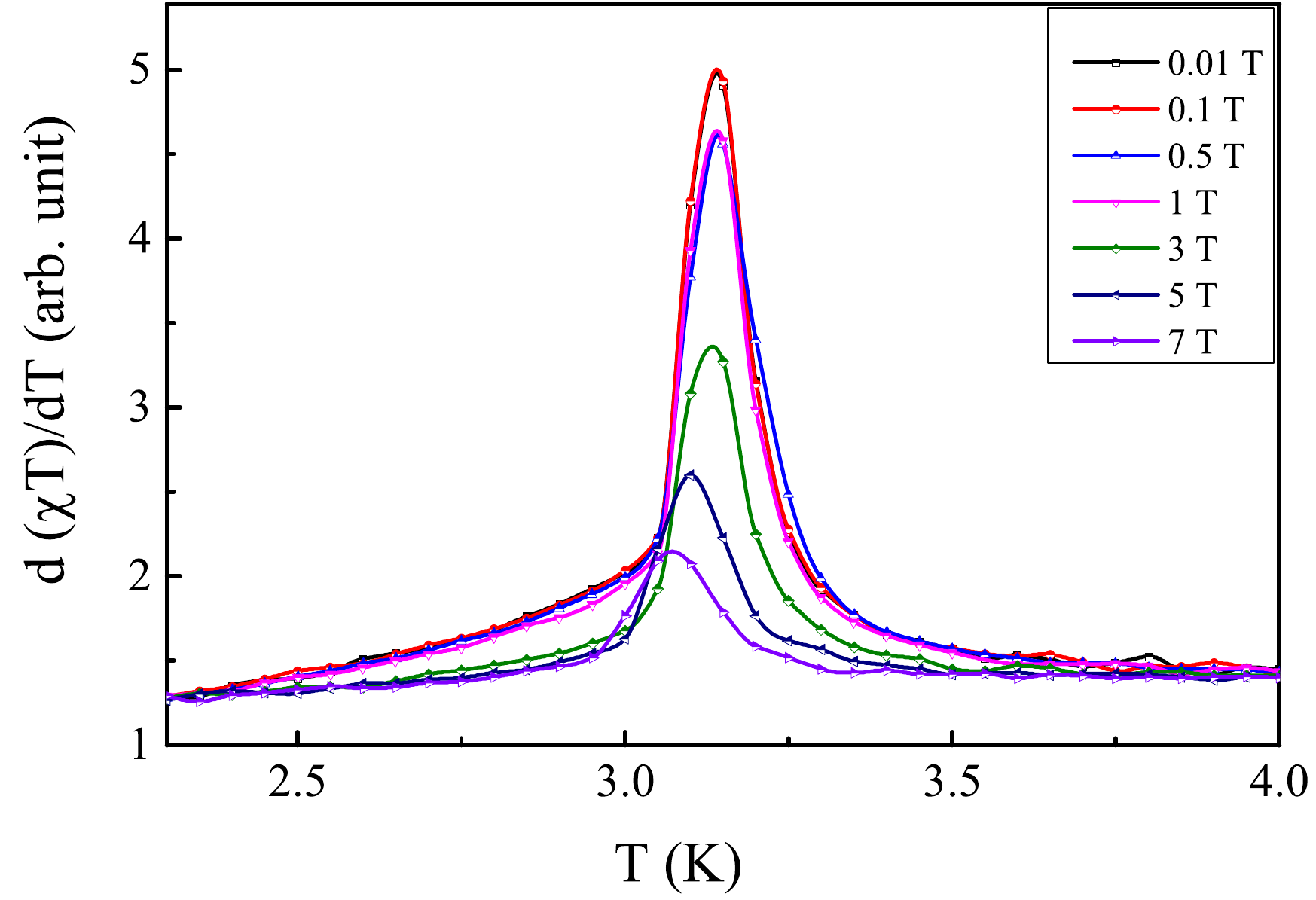}
\caption{
Fisher's heat capacity, $d(\chi T)/dT$, obtained from the susceptibility data measured in different fields. Note that $T_N$ shifts only weakly with increasing the field.
}
\end{figure*}

\begin{table*}
\begin{center}
\begin{minipage}{15cm}
\caption{\label{table} Hoppings and magnetic interactions calculated for different lattice parameters of K$_2$IrCl$_6$ using $U_{\rm eff}=2.2$\,eV, $J_H=0.3$\,eV, and $\lambda=0.4$\,eV, as explained in the main text.}
\begin{ruledtabular}
\begin{tabular}{cccccccc}
$a$ (\AA) & $t_1$ (meV) & $t_2$ (meV) & $t_3$ (meV) & $t_4$ (meV) & $J$ (K) & $K$ (K) & $\Gamma$ (K) \\
\hline
11.37  & 2.77    & 4.60    & $-25.04$  & 0       & 0.84     & 0.47     & 0.17      \\
10.37  & 4.30    & 7.97    & $-52.78$  & 0       & 4.63     & 2.04     & 0.60      \\
10.07  & 4.78    & 8.90    & $-66.28$  & 0       & 7.78     & 3.20     & 0.84      \\
9.77   & 5.42    & 9.61    & $-82.87$  & 0       & 12.70    & 4.99     & 1.13      \\
9.57   & 5.55    & 9.32    & $-96.66$  & 0       & 18.16    & 6.77     & 1.27      \\
9.17   & 6.67    & 7.02    & $-130.06$ & 0       & 34.12    & 12.32    & 1.28     
\end{tabular}
\end{ruledtabular}
\end{minipage}
\end{center}
\end{table*}

\end{document}